# Dieterici gas as a Unified Model for Dark Matter and Dark Energy

C. Sivaram, Kenath Arun and R. Nagaraja

Indian Institute of Astrophysics, Bangalore

**Abstract:** The dominance of dark energy in the universe has necessitated the introduction of a repulsive gravity source to make $q_0$ negative. The models for dark energy range from a simple $\Lambda$ term to quintessence, Chaplygin gas, etc. We look at the possibility of how change of behaviour of missing energy density, from DM to DE, may be determined by the change in the equation of state of a background fluid instead of a form of potential. The question of cosmic acceleration can be discussed within the framework of theories which do not necessarily include scalar fields.

A hundred years ago, in 1910, van der Waals got the Nobel Prize for his work on modifying the perfect gas law, to the realistic behaviour of gases, i.e. assumption of no interactions and point particles underlying the perfect gas description was modified. The constant *'b'* introduces the finite volume for the gas particles and *'a'* describes the strength of the interactions between them. So the modified equation is: [1]

$$P = \frac{RT}{(V-b)} - \frac{a}{V^2} \qquad \ldots (1)$$

Recently the dominance of dark energy in the dynamics of the universe as implied by various observations [2, 3, 4] has necessitated the introduction of dark energy for a repulsive gravity source to make $q_0$ negative [5]. The models for dark energy range from a simple $\Lambda$ term (giving negative pressure) to quintessence, Chaplygin gas, etc. [6, 7, 8]

It was recently suggested that change of behaviour of missing energy density (from DM to DE) may be determined by the change in EOS of a background fluid instead of a form of potential (for example a quintessence potential) thus avoiding fine-tuning problems (inherent in shallow potentials, almost massless fields, radiative corrections, etc.)

Within the framework of FRW, one introduces a Chaplygin gas with equation of state $P = -A/\rho^\alpha$, $\alpha = 1$, $A$ is a positive constant, with the density evolving with scale factor now as:

$$\rho = \sqrt{A + \frac{B}{R^b}} \qquad \ldots (2)$$

This gives simple interpolation between dust dominated phase $\rho = \sqrt{B}\, r^{-3}$ and de Sitter (DE) phase $P = -\rho$ through an intermediate regime described by $P = \rho$.

For $\alpha = 1$, this is the EOS associated with parameterization invariance Nambu-Goto d-brane action in $(d+1, 1)$ space-time [9]. This gives rise to a Galilean invariant Chaplygin gas in $(d, 1)$ space-time. The Chaplygin gas is the only one known to admit a supersymmetric generalisation. (It is interesting to note how an equation formulated to describe transonic flow has become useful in understanding the most dominant aspect of the universe, the dark energy.)

The generalised d-brane action $L = \left[1 - \left(g^{\mu\nu}\theta_{,\mu}\theta_{,\nu}\right)^{\frac{\alpha+1}{2\alpha}}\right]^{\frac{\alpha}{\alpha+1}}$ with $\alpha = 1$ (d-brane) leads to the generalised Chaplygin gas, $P = -A/\rho^\alpha$. Effective equation of state for intermediate regime is given by:

$$\rho = A^{1/(1+\alpha)} + \frac{1}{(1+\alpha)} \frac{B}{A^{\alpha/(1+\alpha)}} R^{-3(1+\alpha)} \qquad \ldots (3)$$

It now appears that a van der Waals type of gas could also play the role of dark energy. We can describe the interaction between dark matter particles through a van der Waals type of gas. [10]

As the DM cloud cools, below a certain temperature, the second term (in equation (1)) will be dominant.

$$P = -\frac{a}{V^2} \Rightarrow P = -a'\rho^2 \qquad \ldots (4)$$

Where, $\rho \propto 1/V$

The equation of continuity in an expanding universe is given by:

$$\dot{\rho} + \frac{3\dot{R}}{R}(\rho + P) = 0 \qquad \ldots (5)$$

$\dot{R}/R$ is the instantaneous rate of expansion.

Using the above expression (equation (4)) for pressure we have:

$$\dot{\rho} + \frac{3\dot{R}}{R}(\rho - a'\rho^2) = 0 \qquad \ldots (6)$$

On solving we get:

$$\rho = \frac{1}{(a' - R^3)} \qquad \ldots (7)$$

If $a' \propto 1/\rho$, the solution corresponds to that of vacuum pressure.

This is similar to the bag model of quarks, as the quarks (within a meson or baryon) get closer together, the force of containment gets weaker so that it asymptotically approaches zero for close confinement. The implication is that the quarks in close confinement are completely free to move about. More the force with which the quarks are pulled apart, the greater the force of containment.

At high enough densities, we can have the possibility where the second term (in equation (6)) dominates, i.e.

$$\rho - a'\rho^2 < 0 \qquad \ldots (8)$$

The equation of continuity becomes:

$$\dot{\rho} + \frac{3\dot{R}}{R}(-a'\rho^2) = 0 \qquad \ldots (9)$$

And on solving we get:

$$R = \exp\left(-\frac{1}{a''\rho}\right) \qquad \ldots (10)$$

The density as a function of $r$ is given by: $\rho(r) = \sqrt{A + B/r^6}$

At small $r$, the second term dominates, giving the usual expression, $\rho = \sqrt{B}/r^3$ and at large $r$, $\rho = \sqrt{A}$

In terms of the volume: 
$$\rho(V) = \sqrt{A + \frac{B}{V^2}} \qquad \ldots (11)$$

And the corresponding pressure: $P(V) = -\dfrac{A}{\sqrt{A + \dfrac{B}{V^2}}}$ ... (12)

The temperature of the expanding volume is given by:

$$T(V) = \dfrac{1}{S\sqrt{AV^2 + B}} \qquad ...(13)$$

$$\dfrac{P}{\rho} = -\dfrac{A}{A + B/V^2} \qquad ...(14)$$

As the volume $V \to \infty$, $\rho \to 1$, $P \to -1$, which is the usual DE dominated universe.

If $T_C$ is the temperature when the transition occurs, then:

$$\rho(T) = \dfrac{\sqrt{A}}{\left(1 - \dfrac{T^2}{T_C^2}\right)^{1/2}} \qquad ...(15)$$

And:

$$P(T) = -\sqrt{A}\left(1 - \dfrac{T^2}{T_C^2}\right)^{1/2} \qquad ...(16)$$

$$\omega(T) = \dfrac{P}{\rho} = -\left(1 - \dfrac{T^2}{T_C^2}\right) \qquad ...(17)$$

For $T \ll T_C$, $P(T) = -\sqrt{A}$, which gives a cosmological constant with negative pressure and $\omega(T) = -1$, which is the characteristic of the present epoch.

The general form of van der Waals equation is given by:

$$P = \dfrac{\alpha \rho c^2}{1 - \beta \rho} - \gamma \rho^2 \qquad ...(18)$$

When $\beta, \gamma = 0$, the equation reduces to the usual expression: $P = \alpha \rho c^2$

$\beta = \dfrac{1}{3}V_{crit}$; $\gamma = 3P_{crit}V_{crit}^2$, where, $V_{crit}$, $P_{crit}$ are the critical volume and pressure respectively

In the equation of state, for negative pressure, we have:

$$\rho + 3P \leq 0 \qquad ...(19)$$

Using the generalised van der Waals equation:

$$\rho\left[1+\frac{3\alpha}{1-\beta\rho}-3\gamma\rho\right]\leq 0 \qquad \text{... (20)}$$

$$\rho=\rho_0\left[a+\frac{(1-a)}{R^{3(a+1)/2}}\right]^2 \qquad \text{... (21)}$$

If $\gamma \propto \frac{1}{\rho^3}$, the equation of state corresponding to that of the Chaplygin's gas: [11]

$$a=\frac{b}{1+a}\frac{1}{\rho_0^{1/2}} \qquad \text{... (22)}$$

When $\beta, \gamma = 0$ and $\alpha = -1$, the equation corresponds to that of the cosmological constant:

$$P=-\rho c^2$$

For the generalised Chaplygin gas we have: [6]

$$\rho=\rho_0\left[a'+\frac{(1-a')}{R^{3(1+b)}}\right]^{1/(1+b)} \qquad \text{... (23)}$$

$$a'=\frac{a}{\rho_0^{(1+b)}}; \quad P=-\frac{A}{\rho^\alpha}$$

The solution for the Chaplygin gas in the equation of state gives:

$$\rho(t)=A+\rho_0 R^{-3} \qquad \text{... (24)}$$

The first term corresponds to DE and the second term to DM. With the expansion of the universe, i.e., with increasing R, the first term (dark energy term) dominates.

The Dieterici equation of state is given by: [12]

$$P=\frac{R_g T}{V-b}\exp\left(-\frac{a}{R_g TV}\right) \qquad \text{... (25)}$$

Where, *a* and *b* have the same meaning as in van der Waals equation.

And the reduced coordinates are: $V_C = 2nb; \quad T_C = \frac{a}{4R_g b}$

In terms of the density, the Dieterici equation of state is given by:

$$P=R_g T\rho\exp\left(-\frac{a}{R_g T}\rho\right) \qquad \text{... (26)}$$

Using this in the continuity equation (equation (5)), we have:

$$\dot{\rho}+(\rho+A\rho\exp(-B\rho))\frac{\dot{R}}{R}=0 \qquad \text{...(27)}$$

Where, $A = R_g T$; $B = \dfrac{a}{R_g T}$

On integration we get the solution as:

$$R = \exp\left(\frac{\rho^2}{2}\right) + \exp\left(\frac{1}{2}A\rho^2\exp(-B\rho)\right) \qquad \text{...(28)}$$

We get a super-exponential term which could be relevant in the early universe, for an alternative inflation.

On expanding the exponential term in the Dieterici equation we have:

$$P = \frac{R_g T}{V-b}\left(1 - \frac{a}{R_g TV} + \left(\frac{a}{R_g TV}\right)^2 ...\right) \qquad \text{...(29)}$$

which reduces to the Van der Waal's equation:

$$P = \frac{R_g T}{V-b} - \frac{a}{V(V-b)} \qquad \text{...(30)}$$

(neglecting higher powers)

For large volume $(V-b) \approx V$, therefore we have:

$$P = \frac{R_g T}{V-b} - \frac{a}{V^2} \qquad \text{...(31)}$$

As $T \to 0$, the second term becomes significant, i.e.:

$$P = -\frac{a}{V^2} \qquad \text{...(32)}$$

The interacting term $\propto -a'\rho^2$

If the interaction between DM particles is non-abelian like QCD, then it would be asymptotically free, i.e., interaction increases with distance.

$a' \propto \dfrac{1}{\rho^2}$ implies $-a'\rho^2 =$ constant = negative pressure = $-B$ (a bag term)

This then becomes similar to a $\Lambda$ term, with negative pressure.

This is similar to the asymptotic behaviour in quantum chromo-dynamics (QCD), where the strength of the interaction increases with the distance, characterised by a bag term.

The bag term in QCD [13] is $\sim \dfrac{m_{QCD}^4 c^3}{\hbar^3} \sim (0.2 GeV)^4$ ... (33)

It is well known that the bag term in QCD is similar to the cosmological term in GR [14], where the bag term has units of $(GeV)^4$.

Here we would have the corresponding term $\propto m_{DM}^4$

It turns out that if dark matter mass is of the axion mass $\sim 10^{-3}$ eV, this indeed gives the right order of negative pressure. [15, 16]

At the Planck epoch $\Lambda_{Pl} \propto M_{Pl}^4$. This suggests a drop by 120 orders, i.e.: [17]

$$\left(\dfrac{m_{DM}}{M_{Pl}}\right)^4 \approx \left(\dfrac{10^{-3}}{10^{28}}\right)^4 \sim 10^{-120} \quad \text{... (34)}$$

At around the Planck or GUTS epoch, such a negative pressure bag term would have driven an exponential inflation expansion.

For heavier DM particles, e.g. 100GeV, a weak interaction coupling amongst the DM particles of $\sim 10^{-42}$ (weaker than gravity) would give the right order of DE. [18]

We can have an acceleration equation of the expansion given by:

$$\dfrac{\ddot{R}}{R} = -\dfrac{4\pi G m^4 c^3}{3 h^3}\left(1 - \dfrac{m_{DM}^2}{m_W^2}\right) \quad \text{... (35)}$$

where the Planck mass associated with the weak interaction between the dark matter particles is given by: $m_W = \left(\dfrac{hc}{G_W}\right)^{1/2}$, where $G_W$ is the Fermi constant. [19]

It is possible, as suggested in string theory as well as in quantum gravity and Kaluza-Klein (KK) theories, that mass spectrum be in multiples of Planck mass, i.e., $m_{DM} > \left(\dfrac{hc}{G}\right)^{1/2}$.

Therefore in this case: $\ddot{R} =$ positive. Therefore the expansion of the universe was repulsive hence giving the dark energy scenario. [20, 21]

Another approach is to consider long range interaction between DM particles to be described (for instance) by a Lennard-Jones type of potential (having both repulsive and attractive interaction):

$$V = I\left[\frac{A}{r^{12}} - \frac{B}{r^6}\right] \quad \ldots (36)$$

Here $I$ would correspond to binding energy.

When applied to liquids the surface tension turns out to be: [22]

$$\sigma = \frac{NRy\beta}{a^2} \quad \ldots (37)$$

Where, $Ry$ is the Rydberg constant and $a$ is the Bohr radius and for their usual values, this works out to be of the order of ~$10^2$ ergs/cm$^2$.

In the case of the DM particles the corresponding value turns out to be $10^{20}$ ergs/cm$^2$, which corresponds to the dark energy density. This would explain various scaling relations of large scale structures, i.e., for structures to form, the gravitational binding energy is of the order of the dark energy density, i.e. background repulsion matches inward gravitational attraction. [23]

$$\frac{GM^2}{8\pi R^4} \approx \frac{\Lambda c^4}{8\pi G}$$

$$\frac{Mc^2}{R^2} \approx \frac{\sqrt{\Lambda}c^4}{G} \approx 10^{20}\, ergs/cm^2 \quad \ldots (38)$$

In conclusion, we see that the change of behaviour of missing energy density may be determined by the change in EOS of a background fluid instead of a form of potential like for instance in the case of a quintessence potential. It is shown how the Dieterici gas model for DM can naturally lead to the various observed phases of the expansion of the universe, including inflation in the early universe and negative pressure of DE as the expansion continues beyond a certain critical size.